\documentstyle[pra,aps,multicol,psfig]{revtex}

\def\addcontentsline#1#2#3{\relax}
\begin{document}
\draft
\title{Features of renormalization induced by interaction 
in 1D transport. }
\author{V.V. Ponomarenko$^{1,*}$ and N. Nagaosa$^2$}
\address{$^1$ Department of Theoretical Physics, University of Geneva,
24, quai Ernest-Ansermet, 1211 Geneva 4 \\
$^2$Department of Applied Physics, University of Tokyo,
Bunkyo-ku, Tokyo 113, Japan}
\date{\today}
\maketitle
\begin{abstract}
One-dimensional interacting electrons in a quantum wire 
connected to reservoirs are studied theoretically.
The difference in the Tomonaga-Luttinger interaction constants
between the wire $(g)$  and reservoirs $(g_{\infty})$ 
produces the cross-correlation between the
right- and left-going chiral components of the charge density wave field. 
The low energy 
asymptotics of this field correlator, which is determined by
$(g)$  and $(g_{\infty})$,  specifies renormalization of 
physical quantities. 
 We have found that charge of the carriers in the shot noise is determined by 
$g_\infty$ (no renormalization for the Fermi liquid reservoirs) at any 
energy, meanwhile the factor $g$ renormalizing 
the charge and spin susceptibilities emerges in the threshold structures at 
some rational fillings. 
\end{abstract}


%
\multicols{2}

\section{Introduction}

Recent development in the nano-fabrication technique makes 
1D interacting electron systems an experimental reality. This allows
comparison of the transport experiments with the predictions 
of the 1D quantum field theory, which has been developing since the early 1930's.
A basic property of this theory suggests that even a weak electron-electron
interaction changes the nature of elementary quasi-particles in
1D. Instead of  free-electron behavior they acquire a fractional
charge and statistics. To deal with the metallic phase of these systems,
a simple model called the Tomonaga-Luttinger liquid (TLL) has been developed.
This model is specified by one constant $g$ of the interaction.
Its deviation from the free-electron value $g=1$ renormalizes 
the density of states of the sound wave excitations,
the compressibility, and the charge of  TLL quasiparticles.
Therefore, it had been expected that the same factor of the fractional charge 
has to appear in the conductance\cite{kf1} 
and in the shot noise \cite{kf2}. 
It had been assumed that this renormalization would
show up in 1D transport through a quantum wire and through the edge of 
a Fractional Quantum Hall Liquid (FQHL). 
First transport measurements have shown that,
in contrast to the above expectation, 
there is { \it no} renormalization of the 1D conductance  of the clean
quantum wire \cite{tar}, although there {\it is}
the renormalization of the conductance and
shot noise in transport through the edge of the  
FQHL \cite{noise}. The first result has been explained by taking into 
 account the Fermi liquid source and drain reservoirs
with the inhomogeneous TLL model (ITLL) of 1D transport \cite{1dcond}.  
Its solution 
clarified that the zero-frequency conductance is unchanged in the TLL wire
at any temperature/voltage and has a simple physical explanation:
The interaction in the TLL wire reduces to forward scattering, and cannot 
change the outgoing flows of electrons from the reservoirs whose chemical potentials are shifted by the applied voltage.
It has been suggested later \cite{chkl} that the same model describes 
transport through the 
FQHL connected with the reservoirs via two point-like contacts. 
However, in typical experiments these contacts are wide. 
Then the equilibration of the edge chemical potential with the reservoir
is expected  \cite{cham} to account for the fractional value of conductance 
equal to the Hall conductivity. 
This suggests  \cite{alekseev} that both problems might be treated in a unified
fashion in terms of an effective voltage which, up to a factor $e$ of electron charge, coincides 
with the difference between the chiral electrochemical potentials. The latter is equal to the real voltage for the FQHL transport according to the above
hypothesis of the equilibration  and has to be rescaled by a factor $g^{-1}$ in the case of the TLL wire \cite{kawa}.  
Therefore, it was claimed \cite{spin} that an exact solution to the 
problem of a point impurity in uniform TLL describes both the FQHL edge
and the TLL wire transport with the above choice of the effective voltage.

In this paper we discuss transport through the 1D quantum wire
in terms of the inhomogeneous TLL (ITLL) model \cite{1dcond,pon2}
to gain a further insight into the role of the reservoirs. Our final
objective is specification of an easy observable parameter in 1D transport 
through the wire which  may confirm  existence of the TLL phase in the latter
when its conductance shows an approximately power suppression \cite{tar,yacobi}
with lowering temperature $ T $. In particular, we 
examine shot noise whose relation to the current doesn't contain 
the applied voltage $V$ and therefore  could along the lines of the above argument, 
reveal a fractional charge of the carriers as it appears in the uniform 
wire solution \cite{fll} addressed to the FQHL transport \cite{kf2}.

This paper is organized as follows. We describe our model in Section II.
For electrons with spin the model assumes 
that the reservoirs are Fermi Liquid, and may be mapped onto the free TLL with the constant $g_\infty =1$,
whereas the wire contains repulsive TLL with $g<1$. Inside the wire a weak backscattering is produced by random impurities, and by Umklapp interaction due to a periodic potential near some rational fillings. 
In the spinless case $g_\infty < 1$ could also occur when  the reservoirs are 
in a principal FQH state with a single edge state along the boundary.

In Section III we derive an expression for the backscattering current 
which is an operator proportional to the rate of particle exchange between 
the right and left chiralities. Its average shows a deviation of the current 
average below its unsuppressed value  $2 e^2 g_\infty V/h$. In our 
derivation this operator is determined by a low frequency asymptotics of the 
retarded correlator of the charge density wave field. The operator specifies 
an energy and charge transferred by the one-particle exchange as 
$e g_\infty V$ and $g_\infty e$,
respectively.  Although this energy neither fixes  the difference between the 
chiral electrochemical potentials unambigiously, nor contradicts their 
above choice, together with the charge they do so. 
Moreover, this charge
has important physical implications by itself, in particular, for the 
current fluctuations. 
We analyze the low frequency limit of these fluctuations in Section IV. 
First, we find a general expression connecting fluctuations of the direct 
and  backscattered currents in the ITLL model. It shows that both fluctuations approach each other 
when  $V \gg T$ and allows us to derive a linear relation 
between fluctuations of the direct current, i.e., shot noise,  and the average of the backscattered current.
For the metallic wire this relation always has $g_\infty$ as the coefficient.
It reveals that the zero-frequency current fluctuations are determined 
by the backscattering of carriers of the charge $g_\infty e$. It is not only 
valid for energies less than that ($T_L$) of the wire length, where the average 
current behavior corresponds to the reservoir type spectrum (the Fermi liquid
behavior at $g_\infty=1$ \cite{prb}), but also for energies above $T_L$. 
At these energies
the current is ruled by the TLL spectrum of the wire. This means, in 
particular, that the above idea \cite{spin,kawa,alekseev} of accounting for 
the reservoir effect by a proper choice of the effective voltage oversimplifies 
the problem and the exact solution by Lesage  {\it et al} can not be addressed 
to the wire transport in the way they suggested \cite{spin}. 

Our further search for an observable parameter, whose renormalization by the
interaction endures connecting the wire to the reservoirs,  
starts from observation that 
the factor $g$ appears in the low frequency asymptotics of the retarded correlator. Then it eventually multiplies the chemical potential of the wire, i.e., the renormalization  of TLL compressibility is not affected by the reservoirs. In Section V we suggest how this factor can be observed in the threshold  structures \cite{us} produced by the Umklapp scattering on the 
periodic potential of the wire in current vs voltage. 
When a filling factor $ \nu $ of electrons in the wire  related 
to the period of the potential is close to its
rational value (1, 1/2, 1/3, etc), this structure 
arises as the current suppression by the Umklapp backscattering is 
strengthened when voltage exceeding an energy $E_{thr}$ proportional to the quasi-momentum is transferred. The resonant Umklapp backscattering at an even
denominator $\nu $ induces charge transfer only, and the coefficient of the
proportionality is $v_c$, the velocity of the charge excitation. Meanwhile 
at odd denominator $\nu$ there are two threshold voltages proportional
to velocities of the charge and spin excitations $v_c$ and $v_s$, respectively. This structure has been observed \cite{kouw} in transport through a 1D wire with the periodic potential induced artificially.
However, the interpretation of the  experiment lacks an understanding of
the interaction effect. Recently Tarucha et al. \cite{tar2} succeeded in 
introducing potential of a shorter period into a more narrow 1D wire. The electron density $\rho_c$ can be continuously controlled by the gate voltage, 
and one can satisfy the half-filling condition 
within an accessible value of $\rho_c$.
Then observation of  a variation of $E_{thr}$ with changing the average chemical potential would probe the TLL inside the wire. 
However, in this section we will see that the method may be used 
only under a severe restriction to the capacitance density
between a close screening gate and the wire. Otherwise the classical electrostatics essentially modifies the TLL compressibility. 
If the density is large enough,  $E_{thr}$ related to the charge transfer shifts as $\Delta E_{thr}=g e V$ under asymmetrically applied voltage
when only one chemical potential of the reservoirs is biased. The threshold voltage related to the spin transfer shifts as
$\Delta E_{thr}=g e V v_s/v_c$  ($e=\hbar=1$  below).
Similarly, under application of a magnetic field $H$,  $E_{thr}$ produced by
the Umklapp processes involving the spin transfer splits into two thresholds divided by a gap equal to $\mu_e H$, with the electron magnetic momentum
$\mu_e$ renormalized by the interaction. Finally, we summarize our 
findings and sketch an attempt at their generalization in Section VI.

\section{Model}

Our model can be derived following \cite{1dcond}
from a 1 channel electron Hamiltonian 
\begin{eqnarray}
\lefteqn{ {\cal H}= \int d\!x \{ \sum_\sigma \psi_\sigma^+(x) ( - \frac{
\partial^2_x}{2m^*} - E_F)\psi_\sigma(x)}
\nonumber\\
& & \hspace{10mm} + \varphi(x) \rho^2(x) +
[V_{imp}(x) + V_{period}(x)] \rho(x) \} ,
\label{x1}
\end{eqnarray}
with the periodic potential $V_{period}(x)$ ( period $a$) 
producing Umklapp backscatterings and the random impurity potential $V_{imp}(x) \rho(x)$.
The Fermi momentum $k_F$ and the Fermi energy $ E_F$
is determined by the filling factor $\nu$ as $\nu=k_F a/\pi$ 
and $ E_F \approx v_F k_F$.
In Eq. (\ref{x1}) the function $\varphi(x)= \theta (x) \theta (L-x)$ switches on the
electron-electron interaction inside the wire confined in $0<x<L$.
Following Haldane's 
generalized bosonization procedure \cite{hald} to account for the nonlinear 
dispersion one has to write the fermionic fields as 
$\psi_{\sigma}(x)= \sqrt{k_F/(2 \pi)} 
\sum exp\{i(n+1)(k_Fx+\phi_{\sigma}(x)/2) +i\theta_{\sigma}(x)/2 \}$
and the electron density fluctuations as $\rho(x)=\sum \rho_{\sigma}(x), \ 
\rho_{\sigma}(x)=(\partial_x \phi_{\sigma}(x))/(2 \pi) 
\sum exp\{in(k_Fx+\phi_{\sigma}(x)/2)\}$
where summation runs over even $n$ and 
$\phi_{\sigma}, \theta_{\sigma} $ are mutually conjugated bosonic fields 
$[\phi_{\sigma}(x), \theta_{\sigma}(y)]=i 2 \pi sgn(x-y)$. 

After substitution of these expressions into (\ref{x1}) and introduction of 
the charge and spin bosonic fields as 
$\phi_{c,s}=(\phi_\uparrow \pm \phi_\downarrow)/ \sqrt{2}$ 
we come to the bosonic form ${\cal H}_B$ of  the Hamiltonian (\ref{x1}). We skip writing it down ( see \cite{us}), as below we will use its associate Lagrangian with respect to the $\phi$-fields. It will appear in our calculation of  the averages of operators which are functions of the $\phi_{c,s}$ fields. Considering 
$1/(4 \pi) \partial_x \theta_{c,s}(x)$ as the conjugated momentum to the field
$\phi_{c,s}(x)$, respectively, one can find the density of the Lagrangian as
\begin{equation}
{\cal L}=\sum_{b=c,s} {\cal L}_b(x,\phi_b,\partial_t \phi_b) + 
{\cal L}_{bs}(x,\phi_c, \phi_s) .
\label{-1}
\end{equation} 
The first part of the Lagrangian describes a
free-electron movement modified by the forward scattering interaction.
Its density is a quadratic form \cite{1dcond}:
${\cal L}_b={1 \over 2} \int d x \phi_b(x) \hat{K}_b \phi_b(x)$, with a differential
operator:
\begin{equation}
\hat{K}_b= - {\partial_t^2 \over {4 \pi v_F} }
 + \partial_x  \left({{v_b(x)}
     \over {4 \pi g_b(x)}} \right) \partial_x .
\label{1}
\end{equation}
Here $v_F$ is the Fermi velocity in the channel, 
$g_c(x)=g$ for $x \in [0,L]$, otherwise  $g_c(x) \equiv g_\infty =1$ ( $g$ is less than 1 for the repulsive interaction and it will be assumed below ), 
and $v_c(x)=v_F/g_c(x)$. The constant of the spin channel 
approaches its low energy value $g_s=1$ fixed by $SU(2)$ symmetry and 
$v_s \le v_c$. 

The second part of the Lagrangian density
specifies \cite{us}
the most slowly decaying terms of backscattering among others with the same
transferred momentum as
\endmulticols
\vspace{-6mm}\noindent\underline{\hspace{87mm}}
\begin{equation}
{\cal L}_{bs}= - \frac{E_F^2 \varphi(x)}{v_F} \bigl[
\sum_{even \  m >  0} U_m \cos(2 k_{mF} m x + 
{{m \phi_c(x)} \over \sqrt{2}}) + \sum_{odd \  m \ge 1} (\delta_{m,1} 
{ V_{imp}(x) \over {2 \pi E_F}} + U_m )
\cos({\phi_s(x) \over \sqrt{2}})
cos(2 k_{mF} m x + {{m \phi_c(x)} \over \sqrt{2}}) \bigr] .
\label{2}
\end{equation}
\noindent\hspace{92mm}\underline{\hspace{87mm}}\vspace{-3mm}
\multicols{2}\noindent
Here $\varphi(x)= \theta (x) \theta (L-x)$ switches on the
backscattering interaction inside the wire confined in $0<x<L$.
The Fermi momentum $k_F$ relates to the Fermi energy $ E_F$ as
$ E_F \approx v_F k_F$, to 
the filling factor $\nu$ as $\nu=k_F a/\pi$, and to 
the transferred momenta $2 k_{mF}$ in the absence of applied voltage:
$k_{mF}=k_F -  \pi l/(m a))>0$, where $l$ is an integer chosen to minimize 
$k_{mF}$.  Depending on the value of $g$ a few terms of the Umklapp
backscattering could open gaps $M$ in the spectrum of the infinite wire
at small $k_{mF}$.  
The most singular is the first term of the second sum ($m=1$) responsible
for opening of the band gap in the infinite wire at $\nu =1$.  The first 
term of the first sum ($m=2$) produces a Hubbard gap at half filling. 
Away from these fillings
all backscattering is produced by the random impurity potential $V_{imp}$
and  the wire is in a metallic phase (TLL) or an Anderson insulator. 
General consideration of the backscattering current and the
current fluctuations in  Sections III and IV does not assume any small
parameters. To probe shot noise, however, we treat the backscattering in
the lowest perturbative order. In Section V we neglect the random potential and assume the perturbative regime of the Umklapp interaction
$M/T_L \ll 1$ which has been observed experimentally \cite{kouw}.

Application of smooth electric ($-\partial_x V(x)$) and magnetic $(H(x))$ fields may be accounted for by addition of  the following Lagrangian density,
\begin{equation}
{\cal L}_V+{\cal L}_H={1 \over {\sqrt{2} \pi}} \{ \phi_c \partial_x V(x)
+{1 \over 2} \phi_s H(x) \},
\label{LV}
\end{equation}
to the one in (\ref{-1}). The function $V(x)$ is equal to the electrochemical potential of the left (right) reservoir $V_L (V_R)$ outside the wire $x<0 (x>L)$. Inside the wire it will be determined self-consistently in the case of  a uniformly screened wire by a close gate in the absence of impurities in Section V. To describe the effect of the uniform magnetic field we put
$H(x)=H$ inside the wire, $0<x<L$, and switch the field off at the infinities
$H(\pm \infty)=0$. 

With a finite voltage applied between the reservoirs $V=V_L-V_R$, 
the system is in non-equilibrium. Then calculation of a  physical quantity average at some moment of time involves consideration of the time evolution of  the system along the Keldysh contour $\Gamma$ which runs from $-\infty$ to $+\infty$ above the real time axis and returns back to $-\infty$ below it. 
In particular,  the current flowing through the point $x$ at the time $t_0$
is given by the average of its operator 
\[ I(V)=<\hat{I}>=-<\partial_t \phi_c(x,t_0)>/(\sqrt{2}\pi) \] .
It  does not depend on $x$ and $t_0$ for the stationary voltage. Describing the evolution in functional integral technique allows us to write it  as 
\begin{eqnarray}
&I&=-\int ( \prod_b D \phi_b ) \frac{\partial_{t_0} \phi_c(x,t_0-i0)}{\sqrt{2}\pi}
e^{i[{\cal S}+{\cal S}_V+{\cal S}_H)]}  
\label{I} \\
&{\cal S}&= \sum_{b=c,s} {\cal S}_b(\phi_b, \partial_t \phi_b)+{\cal S}_{bs}(\phi_c, \phi_s) \nonumber \\
&=&\int dy \int_\Gamma dt [\sum_b {\cal L}_b(y,\{\phi_b,\partial_t \phi_b\})+{\cal L}_{bs}(y,\{\phi_{c,s}\})] \label{S} \\
&{\cal S}&_{V,H}=\int dy \int_\Gamma dt {\cal L}_{V,H} (y, \phi_{c,s})
\label{Sv}
\end{eqnarray}
Here the differential operator acting on the Keldysh time contour in the first  two Gaussian actions ${\cal S}_{c,s}$ in  (\ref{S}) needs further specification. It could be gathered from comparison with the equilibrium model in the absence of   ${\cal S}_{bs}, {\cal S}_{V,H}$. We may write these operators as a $2 \times 2$ matrix operator working on the ordinary
time contour going from $-\infty$ to $\infty$ if vectors $\bar{\phi}_b(x,t), \ (b=c,s)$ are introduced with components $\bar{\phi}_b^T=(\phi_b(x,t_+),\phi_b(x,t_-))$ ($t_{\pm}=t\pm i0$). Then the quadratic Lagrangian density of ${\cal S}_b$ takes its matrix form 
${\cal L}_b={i \over 2} \bar{\phi}_b^T \hat{T}_b^{-1} \bar{\phi}_b$, where the matrix operator $\hat{T}_b$ may be compiled from the equilibrium finite temperature correlators \cite{mah} $T_{b,\pm,\pm}(x,y,t)=<T_{\Gamma}\{\phi_b(x,t_\pm )\phi_b(y,0_\pm )\}>$ in the absence of ${\cal L}_{bs}$ and ${\cal L}_{V,H}$. These correlators are proper solutions to the differential equation related to 
$\hat{K}_b$. They are connected via a spectral density 
representation with the retarded correlator 
$T_{b,R}= T_{b,+,+}-T_{b,+,-}$,
calculated in \cite{pon2}. In general, only Fourier transform of these correlators with respect to $t$ ($\omega$-form) is well determined because of $1/\omega$ singularity. We will use the $t$-form symbolically to make notation shorter.

The above consideration could be easily modified to cover the spinless case,
where $\phi_s$ is absent ($\phi_s=0$) and the charge density is 
$\rho_{c}(x,t)=(\partial_x \phi_{c}(x,t))/(2 \pi)$. The $1/\sqrt{2}$ factor
at the $\phi_c$ field drops out from the backscattering Lagrangian (\ref{2}).
In the spinless case $g_\infty=1/(2n+1)=\nu$ could occur in experiment if the
reservoirs are in a proper FQHL phase. 

\section{Duality transform and backscattering current}

Part of the action ${\cal S}_V+{\cal S}_H $ induced by the external fields tends to shift an average value of $\phi_{c,s}$.  This effect can be accounted for by changing the variables of the functional integration in (\ref{I}),
\begin{equation}
\phi_{b}(x,t_\pm )\equiv \eta_{b}(x,t_\pm )
+\Delta \phi_{b,\pm}(x,t), \ \ b=c,s ,
\label{33}
\end{equation}
so that the charge field satisfies: 
\begin{equation}
{\cal S}_c(\phi_c,\partial_t \phi_c) + {\cal S}_V(\phi_c )=
{\cal S}_c(\eta_c,\partial_t \eta_c)+const.
\label{Seta}
\end{equation}
and the spin field does the same. Let us first consider a non-equilibrium effect of the Lagrangian (\ref{LV}) produced by the voltage. It 
results in such a shift of the charge field, 
\[
\Delta \phi_{c,\pm}(x,t)={i \over {\sqrt{2}\pi}} \int dt' \int dy
T_{c,R}(t-t',x,y) \partial_y V(y) ,
\] 
where the integration over $y$ is, in fact, only inside the wire  $y \in [0, L]$, as $V(y)$ is constant equal to the electrochemical potential of the
left $V_L$ or right $V_R$ reservoir outside it.
The low energy asymptotics of the retarded correlator for $x,y \in [0,L]$
extracted from \cite{pon2} as
\begin{eqnarray}
T_{c,R}(x,y,\omega \approx 0)=2\pi  g_\infty 
({1 \over \omega} +i c(x,y) ), 
\label{21}\\
c(x,y)={t_L \over 2}\left({g_\infty \over g}-{g \over g_\infty}\right)+
\frac{g |y-x|}{v_c g_\infty},
\label{22}
\end{eqnarray}
are specified with two constants $g_\infty, g$ in addition to the 
dimensional and,
hence, non-universal parameters: charge wave velocity in the wire
$v_c$ and traversal time $t_L=L/v_c$. These constants determine the 
coefficients at low $\omega$ and small $x-y$ singularities respectively.
The difference between these two constants $g$ and $g_\infty$ 
indicates the cross correlation between 
the left and right chiral components of $\phi_c$ in the wire.
Remarkably, 
the same constant $g$ in (\ref{22}) also determines the $1/\omega $ behavior 
for high energies  $\omega t_L \gg 1$, and therefore the density of states 
for the charge wave excitations of the TLL inside the wire.

Substituting these asymptotics (\ref{21}), (\ref{22}) one can come to
\endmulticols
\vspace{-6mm}\noindent\underline{\hspace{87mm}}
\begin{equation}
\Delta \phi_{c}(x,t)= \sqrt{2} \biggl[ g_\infty (V(\infty)- V(-\infty))t 
+ \int^{x} dy {g_c(y) \over v_c(y) }(V(+\infty) + V(-\infty) - 2V(y))
\biggr] .
\label{3}
\end{equation}
\noindent\hspace{92mm}\underline{\hspace{87mm}}\vspace{-3mm}
\multicols{2}\noindent
The time dependence of $\Delta \phi_{c}$ relates to the current shift 
under a finite voltage. It feels the entire drop of the potential
$V=V(-\infty)- V(\infty)$ only. The spatial dependence reflects the 
electron density redistribution following the new potential profile.
The shift of the spin field may be written by analogy with (\ref{3}) as
\[
\Delta \phi_s(x)=-\sqrt{2} x g_s H/ v_s
\]
for $x \in [0,L]$. It does not contain a time dependence, since with the choice
of $H(x)$ we have assumed the effect of ${\cal L}_H$ is equilibrium.

Applying the above change of variables (\ref{33}, \ref{3}) under the integral in (\ref{I}), we come to the relation between the average direct and backscattering currents:
\begin{eqnarray}
I&=&\sigma_0 V - \nonumber \\
&\int& ( \prod_b D \eta_b ) \frac{\partial_{t_0} \eta_c(x,t_0-i0)}{\sqrt{2}\pi}
e^{ i(\sum_b{\cal S}_b(\eta_b,\partial_t \eta_b)
+{\cal S}_{bs}(\eta+\Delta \phi))}   .
\label{4}
\end{eqnarray}
Here $\sigma_0$ is the universal conductance $g_\infty e^2/\pi$, and
the second term which we denote $\Delta I$ ensues from the backscattering Lagrangian (\ref{2}). Indeed, this term may be found by expanding 
the exponent under the integral in ${\cal S}_{bs}$ and noticing that the 
gaussian $\eta$-field correlators coincide with the $T$-correlators introduced
in the previous Section.
\endmulticols
\vspace{-6mm}\noindent\underline{\hspace{87mm}}
\begin{equation}
\Delta I = \sum_\pm \frac{-i}{\sqrt{2}\pi} 
\int_{-\infty}^{\infty} dt \int_0^L dy
\partial_{t_0}T_{c,-,\pm}(t_0,t,x,y)\int ( \prod_{b=c,s} D \eta_b ) 
\frac{\delta {\cal S}_{bs}}{\delta \eta_c(t_\pm,y)}
e^{i(\sum_b{\cal S}_b(\eta_b,\partial_t \eta_b)
+{\cal S}_{bs}(\eta+\Delta \phi))}
\label{4a}
\end{equation}
\noindent\hspace{92mm}\underline{\hspace{87mm}}\vspace{-3mm}
\multicols{2}\noindent

Since $ \pm < \delta {\cal S}_{bs}(\{ \eta_b+\Delta \phi_b\}_{b=c,s}/\delta \eta_c(t_\pm,y)))>$ are equal for both signs and do not depend on $t$, the $T$-correlators may be collected into the retarded correlator. Substituting 
its zero frequency asymptotics (\ref{21}), we gather that $\Delta I(t)$ is the average of an operator $\Delta I(t)=<\Delta \hat{I}(t)>$
of the backscattering current,
\begin{equation}
 \Delta \hat{I}(t)\equiv -g_\infty \sqrt{2}
\int dx j(x,t),
\label{bs}
\end{equation}
whose density is 
\begin{equation}
j(x,t_\pm) \equiv \pm
\frac{\delta {\cal S}_{bs}(\{ \eta_b+\Delta \phi_b\}_{b=c,s})}
{\delta \eta_c(t_\pm,x)} .
\label{j}
\end{equation}
Let us relate operator (\ref{bs}) with the one which
counts transitions of  particles between the right
and left chiralities per a unit of time. In Hamiltonian formalism, the latter may be found as
\begin{equation}
\partial_t \hat{N}= {-i\over (2 \sqrt{2} \pi)} \int d\!x \, [\partial_x \theta_c(x), {\cal H}_B]= 
\sqrt{2} \int d\!x {\delta {\cal H}_{bs}(\phi_{c,s}) \over \delta\!\phi_c(x)}, 
\label{N}
\end{equation}
where $ {\cal H}_{bs}(\phi_{c,s})=-\int d\!y {\cal L}_{bs}(y, \phi_{c,s})$, and we use notation ${\cal H}_B$ for the bosonic form of Hamiltonian (\ref{x1}) .
Comparison of (\ref{bs}) with (\ref{N}) proves $\Delta \hat{I}(t)= g_\infty \partial_t \hat{N}$, and that $g_\infty$ in
the coefficient in (\ref{bs}) is the charge of  the particles transferred between the chiralities. It will be corroborated with the shot noise analysis in Section IV. This charge is $g_c(x)$ time less than $e$  as a cloud of opposite charge surrounds each particle. However, the charge 
is taken not at the point of the transition $x$ but at the point of the final
destination $x=\pm\infty$ for each particle, where $g_c(x)=g_\infty(=1)$. 
Substitution of the expression for $\Delta \phi_c$ in (\ref{bs})
reveals that the TLL parameter $g$ enters into the backscattering current
(\ref{bs}), (\ref{j}) in two more ways: (ii) The charge in front of the voltage drop $V(-\infty) - V( \infty)$ in (\ref{3}), (\ref{j})
turns out to be $g_\infty$ too, since both charges and also the 
one in $\sigma_0$ originate from the same parameter of the asymptotics of
the retarded correlator in (\ref{21}); (iii) There appear local values of $g$ unchanged by the reservoirs in the spacious dependence of  $\Delta \phi_c$
in (\ref{3}). We will argue in Section V that they are related to the charge 
compressibility whose values inside the wire are not affected by the reservoirs.

\section{Shot noise}

Next, we consider spectrum of the current fluctuations:
$\delta I^2(V,\omega)=\sum_\pm P(\pm\omega)/2$, where $P(\omega)$ is Fourier
transform of the current-current correlator:
\begin{equation}
P(\omega)=\!\int dt\, e^{i \omega t}(<\hat{I}(t)\hat{I}(0)>-<\hat{I}>^2)=P^*(\omega) .
\end{equation}
Assuming that the current is measured in the right lead,
we find
\begin{equation}
P(t)=
{<\partial_t \eta_c(L,t_-)\partial_t \eta_c(L,0_+)> \over 2 \pi^2}
-<\Delta \hat{I}>^2 .
\end{equation}
Making a similar calculations to those we did for the average current
one can find the spectrum at non-zero $\omega$:
\endmulticols
\vspace{-6mm}\noindent\underline{\hspace{87mm}}
\begin{eqnarray}
P(\omega)&=&{i \omega \over {2 \pi^2}}\partial_t T_{c,-,+}(\omega,L,L)
-{i \over 2 \pi^2} \int dx \langle \frac{\delta^2 {\cal S}_{bs}}
{\delta \eta_c^2(0,x)} \rangle \sum _{\alpha=\pm}\alpha
\partial_t T_{c,-,\alpha}(L,x,\omega) \partial_t T_{c,+,\alpha}(L,x,-\omega)
-{1 \over (2 \pi )^2}\sum_{\alpha, \beta=\pm} \int\!\! \int\!dx\, dx'
\nonumber \\
&\times&
\partial_t T_{c,-,\alpha}(L,x,\omega)\partial_t T_{c,+,\beta}(L,x',-\omega)
\int dt e^{i \omega t} \int (\prod_b D\eta_b) 
\frac{\delta{\cal S}_{bs}}{\delta \eta_c(t_\alpha,x)}
\frac{\delta{\cal S}_{bs}}{\delta \eta_c(t_\beta,x')} 
e^{i(\sum_b{\cal S}_b+{\cal S}_{bs}(\eta+\Delta \phi))} .
\label{5}
\end{eqnarray}
\noindent\hspace{92mm}\underline{\hspace{87mm}}\vspace{-3mm}
\multicols{2}\noindent
The first term describes the current-current correlator in the absence of
the backscattering. It is always of equilibrium and, hence, relates \cite{pon2}
to the frequency dependent conductance of the TLL wire without backscattering
through the fluctuation-dissipation theorem (FDT). 
In the limit
$\omega \rightarrow 0$ the right-hand side of (\ref{5}) 
drastically simplifies after 
applying the low energy asymptotics for the $\partial_t T_{c,\alpha,\beta}$.
They may be found from the spectral representations
\begin{equation}
T_{c,>}(x,y,\omega)=\frac{2\pi \rho_c(x,y,\omega)}
{1-e^{-\omega/T}}= T_{c,<}(x,y,-\omega)
\label{51}
\end{equation}
($<,>$ stand for $+,-$ and $-,+$, respectively), with the spectral density:
\begin{eqnarray}
\rho_c(x,y,\omega)& &={1 \over \pi} Re[T_{c,R}(x,y,\omega)]
\nonumber \\
& & \simeq_{\omega \to 0}{2 g_\infty \over \omega}
(1+O(t_L \omega)^2) .
\end{eqnarray}
Availing ourselves of the low frequency asymptotics of representations (\ref{51})
the zero frequency limit for the second term $(II)$ in (\ref{5}) reduces to
\begin{eqnarray}
(II)& &={i \over 2 \pi^2}\! \int \! dx \langle \frac{\delta j(x,0)}
{\delta \eta_c(0,x)} \rangle 
\partial_t T_{c,>}(L,x,\omega)\sum_\pm \pm \partial_t T_{c,R}(L,x,\mp\omega)
\nonumber \\
& & \to_{\omega \to 0} -8T g_\infty^2\int dx \, c(x) \langle \frac{\delta j(x,0)}
{\delta \eta_c(0,x)} \rangle ,
\label{5a} 
\end{eqnarray}
where $c(x)=c(L,x)$.
Calculation of the third term on the right-hand side of Eq. (\ref{5}) 
involves asymptotics of two others $\phi_c$-correlators:
\begin{eqnarray}
T& &_{c,+,+}(x,y,\omega)=T^*_{c,-,-}(x,y,-\omega)=
\nonumber \\
& &=i Im[T_{c,R}(x,y,\omega)]
+coth\left({\omega \over 2T}\right)Re[T_{c,R}(x,y,\omega)]
\nonumber \\
& &\simeq_{\omega \to 0} 2 \pi i g_\infty c(x,y) 
+{2\pi g_\infty \over \omega} coth\left({\omega \over 2T}\right)(1+O(\omega^2)) .
\end{eqnarray}
These asymptotics contain the $x$-independent parts and the $x$-dependent
ones proportional to $c(x,y)$. Their substitution into the 
third term $(III)$ produces, respectively, the piece 
where integration over $x$ may be accumulated 
into the whole backscattering current and the piece $(X)$ 
where it may not:
\endmulticols
\vspace{-6mm}\noindent\underline{\hspace{87mm}}
\begin{eqnarray}
& &(III)=4T \partial_\omega \bar{P}(\omega)+2 \bar{P}(\omega)+(X) \ 
\mbox{where $\omega \to 0$}
\label{5c}\\
& &(X)=-4iT g_\infty^2\! \int\!\! \int \! dt \,dx\, dy 
[c(y)(\langle  j(x,t_+) j(y,0_+) \rangle 
-\langle  j(x,t_-) j(y,0_+) \rangle ) + 
c(x)(\langle  j(x,t_-) j(y,0_+) \rangle 
-\langle  j(x,t_-) j(y,0_-) \rangle )] .
\label{5b}
\end{eqnarray}
\noindent\hspace{92mm}\underline{\hspace{87mm}}\vspace{-3mm}
\multicols{2}\noindent
Here $\bar{P}(\omega)$ is Fourier transform of the backscattering
current correlator: 
\begin{equation}
\bar{P}(\omega)=\int dt e^{i \omega t}<\Delta \hat{I}(t)\Delta \hat{I}(0)> .
\label{barP}
\end{equation}
The piece $(X)$ plus the second term (\ref{5a}) turn out to be 
zero. It is just a Ward identity for the average density $<j(x,t)>$ of the 
backscattering current following from the time translational
invariance of the action in (\ref{4a}). Indeed, making a change 
of the variable $\eta_c \to \eta_c+cst$ of the functional integration
in the expression for $<j(x,t)>$ one can find that:
\begin{equation}
i \langle {\delta j(x,t) \over \delta \eta_c(t,x)} \rangle 
= \int_{-\infty}^{\infty}\! dt'\! \int dx' \sum_{\pm}
<T_{\Gamma}\{j(x,t) j(x',t'_{\pm})\}> .
\end{equation}
The derivative in (\ref{5c}) is equal to deviation of the differential
conductance from the maximum value $\sigma_0$ of the linear bias conductance:
\begin{equation}
\partial_\omega \bar{P}(\omega)|_{\omega \to 0} = - \partial_V <\Delta I(V)>.
\label{deriv}
\end{equation}
Then the relation between the noise of the direct cirrent $\delta I^2(V,0)$ 
and the backscattering one 
$\delta (\Delta I)^2(V,0)=\bar{P}(\omega)|_{\omega \to 0}$ 
at zero frequency reduces to
\begin{equation}
\delta I^2(V,0)
=2 T \sigma_0 + 4 T \frac{\partial <\Delta \hat{I}> }{\partial V}
+\delta (\Delta I)^2(V,0) .
\label{5'}
\end{equation}
This relation has been earlier derived \cite{fll} for 
the backscattering in the uniform TLL.
It meets  the fluctuation-dissipation theorem in 
the equilibrium limit $V \ll T$. Indeed making use of (\ref{deriv}), we can write the third term on the right-hand side of (\ref{5'}) as
\[
\coth({\omega \over 2T})[\bar{P}(\omega)- \bar{P}(-\omega)]/2|_{\omega \to 0}=
-2T \partial_V <\Delta I(V)> .
\]
Hence, it cancels  a half of the second term and the rest is equal to
$2T I/V, \ V \to 0$. On the shot noise 
side $V > T \to 0$ the two first terms on the right-hand side of (\ref{5'}) are going to zero and zero-frequency fluctuations of the current become equal to the fluctuations of the backscattering one. 

We further implement this result to check on the fractional charge appearance in the shot noise predicted for a uniform TLL \cite{kf2,fll}. It suffices to relate an average of the backscattering current produced by $V_{imp}$ in (\ref{2}) to its fluctuations in the weak backscattering limit, as the Umklapp processes are not relevant for the TLL phase of the wire. In lowest order,
\begin{equation}
<\Delta \hat{I }>=-i\int_{-\infty}^0\!\!dt <[\Delta \hat{I}(0), {\cal H}_{bs}(\{\eta_b(t)+\Delta \phi_b(t) \}_{b=c,s} ) ]>_0 ,
\label{firstorder}
\end{equation}
where $<..>_0$ stands for averaging with the equilibrium Gaussian action (\ref{Seta}) for $\eta_b$. Availing ourselves of the relation between the backscattering current  and the 
backscattering Lagrangian (\ref{bs}), (\ref{j}) and noting that the correlators (\ref{barP}) satisfy  $\bar{P}(\omega)=\exp(\omega/T) \bar{P}(-\omega)$ with $\omega=V$ in equilibrium, one can derive, from (\ref{firstorder}):
\begin{equation}
\delta (\Delta I)^2=-g_\infty \coth(g_\infty V/(2T))<\Delta \hat{I}> 
\label{shotnoise} 
\end{equation}
Its substitution into (\ref{5'}) proves that $\delta I^2=-g_\infty <\Delta \hat{I}>$ for $V \gg T$, i.e.,
the charge of the carriers showing up in the current 
fluctuations Eq.(\ref{shotnoise}) determined by the impurity scattering is the one of the reservoir quasi-particle $g_\infty$.  This result
holds on at any energy while $T \ll V$. At low energy (less than $T_L$) the 
average current $<\Delta \hat{I}> $ follows  a power law of the voltage determined by the reservoirs \cite{prb}. In particular, for the Fermi liquid reservoirs: $<\Delta I>/V \approx -(T_L/E_F)^{(g-1)}L/
(\tau_{sc}v_c)$ if the random impurity
potential $<V_{imp}(x) V_{imp}(y)>={v_F \over \tau_{sc}} \delta(x-y)$ is weak $\tau_{sc} E_F \gg 1$. Meanwhile at the high energy the leading $V^g$ power comes from the wire TLL spectrum. Therefore, the above renormalization of the shot noise charge cannot deny that the transport is
carried by the TLL quasi-particles of the wire. It just shows that the transformation of the wire quasi-particles into the reservoir ones
is extremely robust at the low frequency.  

Approaching an insulating phase of the wire some $m$-Umklapp process could be resonantly strengthened if $k_{mF}L < 1$ \cite{us}. In general we 
represent 
$\Delta \hat{I}=\Delta \hat{I}_{imp}+\sum_{m\ge 1}\Delta \hat{I}_m$ in
obvious correspondence to (\ref{2}). Substitution of this into (\ref{barP}) and
(\ref{firstorder}) shows that  each part of 
$\Delta \hat{I}$ brings additive contribution to both the average
current and current noise, which are related by
\begin{eqnarray}
\delta (\Delta I_{imp})^2=-g_\infty coth(g_\infty V/(2T))<\Delta \hat{I}_{imp}>  , \label{shotnoise2}\\
\delta (\Delta I_m)^2=-mg_\infty coth(m g_\infty V/(2T))<\Delta \hat{I}_m> .\nonumber 
\end{eqnarray}
Then the zero-temperature relation $\delta I^2/| <\Delta \hat{I}>|$ between the current fluctuations and the backscattering average is larger than
$g_\infty$ for $V>T_L$. Eventually when $V \ll T_L$ the one-electron backscattering dominates and this relation decreases to $g_\infty$, as 
$<\Delta I_m>$ is proportional to $ - V^{m^2-1}$ for the even $m$'s and 
$ - V^{m^2}$ for the odd $m$'s.

\section{Renormalized charge/spin compressibility}

In this section we consider variation of the charge density in the wire under a change of the reservoir electrochemical potentials neglecting  any backscattering. The local density is determined by the operator 
$ \partial_x \phi_c(x,t_0)/(\sqrt{2}\pi) $. Calculation of a variation of its average results in
\begin{eqnarray}
\Delta \rho_c(x)&=&\int ( \prod_b D \phi_b ) \frac{\partial_x \phi_c(x,t-i0)}{\sqrt{2}\pi}
e^{i[\sum_{a=c,s}{\cal S}_a+{\cal S}_V+{\cal S}_H)]} \nonumber \\
&=&\frac{\partial_x  \Delta \phi_c(x,t)}{\sqrt{2}\pi}
\label{Drho}
\end{eqnarray}
after making change of the variables of the integration (\ref{33}) in the first 
line of (\ref{Drho}). The notation for the actions was written down in (\ref{S}).
Keeping in mind such a setup where the long wire is uniformly screened by
a close gate we suggest that $V(x)$ equal to $V_R (V_L)$ for $x>L (x<0)$,
respectively, is constant  $V(x)=V_{bott}$ inside the wire \cite{pon2}. 
Then the variation of 
the electron density (\ref{Drho}) is also constant in the wire:
\begin{equation}
\Delta \rho_c = {1 \over \sqrt{2} \pi} \partial_x \Delta \phi_c=
\frac{ g \left( \sum_{a=R,L} V_a-2 V_{bott} \right) }{\pi v_c} .
\label{3'}
\end{equation}
From Eq. (\ref{3'}) one can gather that
variation of the average chemical potential $\mu$ inside the wire under
the applied voltage is
\begin{equation}
\mu=(V_R+V_L)/2-V_{bott} .
\label{mu}
\end{equation}
Indeed, the energy density accumulated inside the 
wire under a constant shift of the charge density $\Delta \rho$ is 
$E_{wire}/L=(\Delta \rho)^2 \pi v_c/(4g)$ according to (\ref{1}), and
\begin{equation}
 \frac{\partial \mu}{\partial \rho_c}=
{ 1 \over L} \frac{\partial^2 E_{wire}}{\partial \rho_c^2}= {\pi v_c \over 2g} .
\label{3''}
\end{equation}
The above expression for the average chemical potential contrasts
with the one for the difference between the chiral chemical potentials. 
The latter emerges if the current 
$I = - { 1 \over {\sqrt{2} \pi} } { \partial \over { \partial t} }
\Delta \phi_c$ related to $\Delta \phi_{c}$ in 
(\ref{3}) is represented as a 
difference between the chiral density variations, i.e., 
$I = v_c( \rho_R - \rho_L)$. Then the difference
between the chiral
chemical potentials can be found using (\ref{3''}) as follows:
\[ \Delta \mu ={g_\infty \over g}(V_L-V_R) \]
This quantity however lacks direct physical sense in the situation where chiralities do not conserve.
In particular,  its matching with the energy transferred by the backscattering current (\ref{bs}) per one
particle implies that the particle carries charge $g$ but not $g_\infty$.

The average chemical potential (\ref{mu}) includes an obscure electrostatic potential $V_{bott}$. In experiment, instead of $V_{bott}$, one could
measure potential $V_g$ of the screening gate having a capacitance $C_g$
with respect to the wire proportional to the wire length: $C_g=c_g L$.
To determine $V_{bott}$  and the charge density in the wire as a 
function of $V_g, V_R,$ and  $V_L$, we will
consider electrostatics of the whole set-up self-consistently.
If the voltage 
is applied symmetrically, $ V_R=-V_L=- V/2$, there is no re-distribution
of the charge and $V_{bott}=0$. 
A non-zero variation of $V_R+V_L$, on the other hand,  changes the average chemical potential and, hence, the density in the wire. The additional charge of the wire $Q$ has to minimize the entire energy
\[
E=E_{wire}(Q)+Q^2/(2C_g)+Q V_g-(V_R+V_L)Q/2 ,
\] 
consisting of the electrostatic one
and an internal energy of the wire $E_{wire}(Q)$. To evaluate the 
$L$-
proportional part of the latter, we 
can use its equilibrium form because of the translational symmetry. 
Therefore, the charge density variation meets
\begin{equation}
\Delta \rho_c=c_{eff} \Delta ({{V_R+V_L} \over 2} - V_g) ,
\label{7-}
\end{equation}
with the density
$c_{eff}=C_{eff}/L$ of the effective capacitance:
\[
C^{-1}_{eff}=C^{-1}_g+(\partial Q/ \partial \mu)^{-1} .
\]
Here $Q(\mu)/L$ is dependence of the charge density on the chemical
potential for the uniformly interacting 1D electrons:
$ \mu(Q/L)=L^{-1} \partial_\rho E_{wire}$. A similar expression for the 
effective capacitance has been derived by B\"{u}ttiker {\it et al} \cite{butt}
in the free electron case
when the last term accounts for the final density of the electron states 
inside the conductor \cite{compr}. 
In the TLL model (\ref{1}) the above derivative 
$\frac{\partial \mu}{\partial \rho_c}$ has been found in (\ref{3''}).
Its substitution brings out the density of the effective capacitance as
follows:
\begin{equation}
c_{eff}=\frac{\frac{\partial \rho_c}{\partial \mu}}
{1+{1 \over c_g}\frac{\partial \rho_c}{\partial \mu}}=
\frac{\frac{2g}{\pi v_c}}{1+\frac{2g}{c_g \pi v_c}} .
\label{7}
\end{equation}
This expression shows that in the experimentally relevant dependence of 
$\Delta \rho_c$ on $(V_R+V_L/2)$ at fixed $V_g$ the TLL compressibility of the wire is renormaliized by a factor $(1+2g/(c_g \pi v_c))^{-1}$ due to
electrostatic Coulomb screening. In the limit: $2 e^2 L/(\pi C_g v_F) \ll 1$
( without 2 for the spinless electrons ), this factor becomes unimportant and $V_{bott}$ in (\ref{3'}) remains independent of $V_R$ and $V_L$.  On the other hand, a finite density of the gate capacitance veils the short range interaction effect on the charge density. 
Let us estimate the gate capacitance contribution for a GaAs wire. 
Evaluating $c^{-1}_g \approx d/(\varepsilon R)$ if the radius of the wire $R$
is larger than the distance $d$ from the wire to the gate, the dielectric
constant $\varepsilon=12$, $v_F=2 \times 10^5 m/s$ and 
$e^2/h=3.3 \times 10^5 m/s$, we
come to $2 e^2 L/(\pi C_g v_F) \approx 0.2 d/R$. This shows that the effect of the finite gate capacitance may be suppressed in  experiment. Then a direct
measurement of the charge density variation, which is equal to minus the variation
of the density in the screening gate, would us allow to find a ratio of the TLL
parameters: $g/v_c$. 

To find the constant $g$, however, the Umklapp backscattering process in (\ref{2}), dependent on the transferred momentum, turns out to be useful. 
Its manifestation in  current vs voltage was described  \cite{us} in the lowest order in the strength of the backscattering for $V_L=-V_R$ as the appearance of a threshold structure. In a general case the charge density variation specified by (\ref{7-}) will show up in this effect causing a shift of this structure.
Indeed, the average Umklapp backscattering current $\Delta\! I$ is  
decomposed into the sum of the different backscattering mechanism contributions
$<\Delta \! I_m>$ in the lowest order. Substituting (\ref{2}) into (\ref{firstorder}), one can find that the even $m$ terms involving only $\phi_c$ field are equal to 
\endmulticols
\vspace{-7mm}\noindent\underline{\hspace{87mm}}
\begin{eqnarray}
<\Delta\! I_m>= -{m \over 4} \bigl({{U_m E_F^2} \over v_F}\bigr)^2
\int_{- \infty}^{\infty}\!dt \int\!\! \int_{0}^{L}\! d \!x_1 d \! x_2
<e^{im \phi_c(x_1,t)/\sqrt{2}} e^{-im \phi_c(x_2,0)/\sqrt{2}}>
\bigl[e^{im(2(k_{mF}+\Delta k_F)(x_1-x_2)+g_\infty Vt)} - h.c. \bigr] ,
\label{64}
\end{eqnarray}
\noindent\hspace{92mm}\underline{\hspace{87mm}}\vspace{-3mm}
\multicols{2}\noindent
where $\Delta k_F=\partial_x \Delta \phi_c/(2\sqrt{2})$
is identified with a variation of the Fermi momentum  $k_{F}$  in agreement 
with the Luttinger theorem in the TLL:$\Delta \rho_c=2 \Delta k_F/\pi$. 
The odd $m$ terms additionally include a spin field
correlator $<e^{i \phi_s(x_1,t)/\sqrt{2}} e^{-i \phi_s(x_2,0)/\sqrt{2}}>$
under the integrals in (\ref{64}). This correlator is chiral: It is a product
of two functions, which are negative powers of their arguments: 
$(x_1-x_2)/v_s \pm t $, respectively, if the latters are less than $1/T$.
Similarly, the charge correlator in (\ref{64}) approaches its chiral high 
temperature asymptotics depending on $(x_1-x_2)/v_c \pm t $ at $T>T_L$, 
whereas the interference stemming from the boundary
scattering makes it more complex at the low temperature. Therefore, the 
$L$-proportional
part of $<\Delta\! I_m>$ in (\ref{64}) shows a singular dependence 
\cite{us}
on its arguments, $E_{thr,c}-g_\infty V$ and $E_{thr,s}-g_\infty V$, smeared over the $max\{T,T_L\}$ range near their zeros if $m$ is odd. Here the charge and spin threshold 
energies are $E_{thr,c}=v_c(2 k_{mF}+2 \Delta k_F)$
and $E_{thr,s}=v_s(2 k_{mF}+2  \Delta k_F) < E_{thr,c}$, respectively.  In particular, suppression of the differential conductance produced by $\Delta I_1$
has peaks at $g_\infty V=E_{thr,s} $and $g_\infty V=E_{thr,c}$, 
which are precursors of the band gap at 
integer filling, whereas the suppressive contribution to the conductance vanishes 
as $([k_{F}+\Delta k_F]L)^{-2(1+rg)}$ if $[k_{F}+\Delta k_F]L \gg 1$ 
at low voltage \cite{us}. 
Similarly, $\Delta I_2$ brings out a peak in the differential 
conductance versus voltage 
at $g_\infty V=E_{thr,c}$ near half filling, which is a
precursor of the Mott-Hubbard insulator. We see that the positions of these
structures vary with the chemical potential inside the wire as
\begin{equation}
\Delta E_{thr,c}=\pi v_c c_{eff} 
\Delta ({{V_R+V_L} \over 2} - V_g)= {v_c \over v_s} \Delta E_{thr,s} ,
\label{68}
\end{equation}
with $c_{eff}$ written in (\ref{7}). Measurement of this variation 
allows us to find $g/g_\infty$
and $v_c/v_s$ if the density of capacitance $c_g$ between the wire and
the gate located closely is large enough. 

Finally, we discuss effect of the magnetic field on the threshold structure.
The linear in $x$ shift $\Delta \phi_s$ produced by this field changes 
momentum transferred by the odd $m$ terms of ${\cal L}_{bs}$ in (\ref{64}).  It results in a symmetrical split of  both of the threshold energies 
$E_{thr,b}\pm \Delta_H E_{thr,b}/2$ associated
with the m-th threshold structure. There is no additional redistribution
of the charge and $\Delta_H E_{thr,c}=v_c g_s H/v_s$
and $\Delta_H E_{thr,s}= g_s H$. The even $m$ threshold energies are 
not affected by the magnetic field.
In the case of the spin $SU(2)$ symmetry,
the charge threshold splitting shows renormalization
of the electron magnetic momentum in $v_c/v_s \approx g^{-1}$ times even though there
is no renormalization of the magnetic susceptibility $g_s=1$. Albeit, 
there could be a deviation
of $g_s$ $\propto (ln[E_F/max\{T_L,T,V\}])^{-1}$ from 
its renormalization group limit \cite{em}.

\section{Conclusion}

In the model we have analyzed, the low energy form of the retarded 
correlator is basically determined by the  two 
Tomohaga-Luttinger interaction constants $g$ in the wire and $g_{\infty}$
in the reservoirs: $g_\infty$ is a
coefficient at the low energy $1/\omega$ singularity, and $g$ at the high
momentum $1/k^2$ singularity. The latter also determines the density wave
spectrum at energies higher than $1/t_L\equiv T_L$ and, hence, 
the scaling powers at these energies. Then we saw
that the same constant $g_\infty$ determines the normalization of the conductance, the
low energy charge in the shot noise, and the charge at the voltage drop.
All of these are equal, therefore, in this model. 
In the TLL phase of the wire the same charge $g_{\infty}$
appears in the shot noise at  high energy, whereas  the 
scaling of the backscattering current unambiguously shows that the transport is carried by the intrinsic excitation of the TLL wire. This is because the mechanism of transformation of the quasi-particles of the reservoirs into those of the wire reduces to an interference of the charge density waves scattered
on the inhomogeneities of the interaction \cite{1dcond,pon2}. 
It is classical by its nature and therefore robust, as opposed to the
case of the wire in the correlated insulator phase.

Let us assume that the above form of the
low energy asymptotics holds for the wider variety of models 
describing transport between the reservoirs through a 1D TLL, even though
this could be difficult to prove in the absence of exact solutions. Then we could conclude that in these models charges in the coefficients at the
conductance, the voltage drop, and in the shot noise are the same,
whereas the charge density compressibility and the scaling powers are determined
by a different constant. This conclusion seems to comply with known
results of experiments on transport through a principal FQHL. 
Indeed, normalization of the conductance and the charge in the 
shot noise have been found \cite{noise} to be equal to each
other and to the Hall conductivity for a FQHL with $\nu=g_\infty=1/3$. 
On the other hand, measurement of the tunneling conductance \cite{chang}
have found a power law scaling with the exponent related to a 
different $g:\ 1/g=2.6$.

\section{Acknowledgments}

The authors acknowledge H. Fukuyama and M. Ogata for useful discussions.
One of us (V.P.) endured a criticism of M. Buttiker we benefitted from
during the 1996 Curacao workshop.
This work was supported by the Center of Excellence at the Japanese Society for Promotion of Science and partly by the Swiss National Science Foundation.

\vspace {1cm}

$^*$ On leave of absence from A.F.Ioffe Physical Technical Institute,
194021, St. Petersburg, Russia; Present address: Department of Physics,
SUNY at Stony Brook, Stony Brook, NY 11794.


\begin{references}
\bibitem{compr} 
In general, this term has to be associated with the 
compressibility
\[
-\frac{1}{L}\frac{\partial L}{\partial P}= {1 \over L}
\left(\frac{\partial^2 E_{wire}}{(\partial L)^2} \right)^{-1}=
\left(\frac{\partial \mu}{\partial \rho_c} \rho^2_c \right)^{-1}=
\frac{\partial \rho_c}{\partial \mu}
\left({\pi \over 2 k_F}\right)^{2} ,
\] 
where $P$ is the pressure in a 1D system of length $L$.
\end{references}
\end{document}